\documentclass[12pt]{iopart}
\usepackage{epsfig}
\usepackage{subfigure}
\usepackage{iopams}
\usepackage{array}
\newcommand{\PTs}{$\mathcal{PT}$-symmetric }%
\newcommand{\PT}{$\mathcal{PT}$-symmetry }%
\eqnobysec
\begin{document}
    \title{The band spectrum of periodic potentials with \PT}%
    \author{Jos\'e M. Cerver\'o and Alberto Rodr\'{\i}guez}%
    \address{F\'{\i}sica Te\'orica, Facultad de Ciencias, Universidad de
Salamanca,\\ 37008 Salamanca, Spain}%
    \eads{cervero@usal.es}%
\begin{abstract}%
    A \textbf{real} band condition is shown to exist for one dimensional periodic \textbf{complex}
non-hermitian  potentials exhibiting  $\mathcal{PT}$-symmetry. We use an exactly
solvable ultralocal periodic potential to obtain the band structure and discuss some spectral features
of the model,  specially those concerning the  role of the imaginary parameters of the couplings.
Analytical results as well as some numerical examples are provided.
\end{abstract}%
\submitto{\JPA}%
\maketitle
%%%%%%%%%%%%%%%%%%%%%%%%%%%%%%%%%%%%%%%%%%%%%%%%%%%%%%%%%%%%%%%%%%%%%%%%%%%%%
\section{Introduction}%
    In a previous series of papers \cite{CERROD1}-\cite{CERROD2}, we have
presented the exact diagonalization of the hermitian Schr\"odinger operator
corresponding to a periodic potential composed of $N$ atoms modelled by delta functions with
different couplings for arbitrary $N$.
This basic structure can repeat itself an infinite number of times
giving rise to a periodic structure
representing a $N$-species one dimensional infinite chain of atoms. To
our surprise this model is exactly solvable and a far from straightforward calculation leads
to an exact band condition. Due to the factorizable form of the solution one has not only the advantage of
closed form expressions but one can also perform computer calculations with an exceedingly degree of accuracy
due to the exact nature of the solution itself. The main physical motivation was indeed to modelate the
band structure of a one dimensional quantum wire. The model can be extended with a minimum
amount of effort to the non-hermitian but \PTs \cite{BENBOET} quantum hamiltonian  and the band
condition becomes real although
the potential is obviously a complex one. The imaginary parts of the
couplings play a central role in defining the band structure which can be modeled at will by
varying this parameter. The purpose of this paper is to present this new solution in a detailed fashion. In a
previous paper a preliminary analysis of the existence and properties of this exact band
model solution were given \cite{JMC2}.
Here we want to present a careful study of the solution in a more rigurous
manner paying due attention to the exciting properties that this rich structure seems to
exhibit.

The idea of performing band structure calculations by using a non-hermitian
Hamiltonian is extremely
promissing  as the examples of \PTs quantum hamiltonians so
far existing in the literature rely
more in aspects concerning the discrete spectrum and bound states 
(\cite{AKUS}, \cite{JMC1}, \cite{LEZNOJ} and \cite{MZNOJ}) and are also far from an actual physical
application. It is not the aim of the present paper to provide a full discussion of the manifold  aspects  of
the \PT in quantum
mechanics. We address the interest reader to the original reference
\cite{BENBOET}, some recent interesting theoretical criticism \cite{MOSTAF} and also to some work done recently in
one dimensional models \cite{NKR}
which may help to understand better the role of \PT in the
framework of one dimensional systems.
In all these papers the imaginary part of the coupling plays a substantial
role in defining the spectrum of bound states. We shall show that in the case of the band structure hereby
presented this imaginary parts are of primary importance in defining the transport properties of the one
dimensional quantum chain.

The plan of the paper is the following. In Section \ref{sec:model} we shall be discussing
the main features of the model just
as a simple model for a quantum wire with only real couplings. Furthermore
the presence of \PT
and the role of the complex couplings will be introduced. The band spectrum
remains analytical and real in
spite of the complex non-hermitian nature of the potential. In Section \ref{sec:spectrum} a
throughout analysis of the band
condition is made in order to clarify the effect of the imaginary parts of
the couplings on the band spectrum. This analysis can be expressed  analytically for some cases where $N$ is not too
large. Otherwise the equations appear to be intractable. Even in this case numerical calculations can
always be performed with a large degree of accuracy.
In section \ref{sec:physics} we make some comments on the band spectrum and
observe the form of the \PTs electronic states.
Several pathologies plaguing the band spectrum for the complex coupling case are
observed. It can be avoided working on a small range of energies for
certain values of the imaginary couplings. The localization of the wave
function over the primitive cell is also found to be altered manipulating only the
imaginary parameters. We close with a section of Conclusions.

%%%%%%%%%%%%%%%%%%%%%%%%%%%%%%%%%%%%%%%%%%%%%%%%%%%%%%%%%%%%%%%%%%%%%%%%%%%%%%%%%%%%%%%%
\section {The model and its band spectrum}
\label{sec:model}
Let us begin with a brief remainder of the solution presented in
\cite{CERROD1} corresponding to a infinite
periodic potential composed of  a basic structure made out of a {\it
finite} number $N$ of equally spaced
deltas with different {\it real} $N$ couplings, repeating itself an
{\it infinite} number of times. If the
spacing is $a$ and
\begin{equation}
    a_j = \frac{\hbar^2}{m e_j^2}
\end{equation}
is the length associated to each coupling ($e_j^2$), the band structure can be
written as
\begin{equation}
    \cos(N Qa) = \mathcal{B}(\varepsilon; a_1, a_2,\ldots,a_N)
    \label{ec:defiban}
\end{equation}
where Q is an arbitrary real number which varies between $\frac{\pi}{Na}$
and $-\frac{\pi}{Na}$ and 
\begin{eqnarray}
    \fl\mathcal{B}(\varepsilon; a_1,\ldots,a_N)=&
    2^{N-1}\sum_{P}h_i...(N)...h_k -2^{N-3}\sum_{P}h_i...(N-2)...h_k\nonumber\\
    &+ 2^{N-5}\sum_{P}h_i...(N-4)...h_k -\ldots
    (-1)^{\frac{N}{2}-1}2\sum_{P}h_i...(2)...h_k \nonumber\\
    &+ (-1)^\frac{N}{2} \label{ec:bpar}\\
    \fl\mathcal{B}(\varepsilon; a_1,\ldots,a_N)=&
    2^{N-1}\sum_{P}h_i...(N)...h_k -2^{N-3}\sum_{P}h_i...(N-2)...h_k\nonumber\\
    &+ 2^{N-5}\sum_{P}h_i...(N-4)...h_k -\ldots
    (-1)^\frac{N-3}{2}2^2\sum_{P}h_i...(3)...h_k \nonumber\\
    &+(-1)^\frac{N-1}{2}(h_1+h_2+\ldots +h_N)
    \label{ec:bimpar}
\end{eqnarray}
for even and odd $N$ repectively.

The symbol $\sum_{P}h_i...(M)...h_k$ means a sum over {\it all products of
M different $h_i$'s with the
following rule for each product: the indices must follow an increasing
order and to an odd index must
always follow an even index and reciprocally}.

The functions $h_j$ have the universal form:
\begin{equation}
    h_j(\varepsilon) = \cos(\varepsilon) +
    \left(\frac{a}{a_j}\right)\frac{\sin(\varepsilon)}
    {\varepsilon}
\end{equation}
and the independent variable is a function of the energy (i.e. $\varepsilon=ka$).
In order to see that this
condition looks much simpler than one might think at the beginning of the
calculation let us list below,
for the benefit of the reader, the first three conditions for the cases
$N=2,\,3$ and $4$.
\begin{eqnarray}
    \cos(2Qa) = 2h_1h_2 -1 \label{ec:cos1}\\
    \cos(3Qa) = 4h_1h_2h_3-(h_1+h_2+h_3) \\
    \cos(4Qa) = 8h_1h_2h_3h_4  - 2 (h_1h_2  +h_1h_4 + h_2h_3 +h_3h_4) + 1 \label{ec:cos3}
\end{eqnarray}
It does not require too much time to write down the band conditions for
fairly large $N$, but more important
is the fact that the {\it exact} formulae (\ref{ec:bpar}) and (\ref{ec:bimpar}) are in itself quite easy to
program for sequential calculations. Using
the band condition we have been able to generate not only a full band
spectrum but also to calculate the density of
states for periodic one dimensional lattices with a high degree of accuracy
(See references \cite{CERROD1} and
\cite{CERROD2}).

The next step is to include \PT in this model. It is not
hard to see that \eref{ec:defiban} is
still {\it real} if we include the following changes:
\begin{itemize}

\item Promote the couplings from {\it real} to {\it
complex}. i.e. $\left(\frac{a}{a_j}\right) \rightarrow r_j + \rmi s_j$

\item Order the potential in a  $\mathcal{PT}$-invariant form, that allow
us to choose a \PTs primitive cell. This
leads to the following
identifications:
\begin{eqnarray*}
    h_N &= h_1^* \\
    h_{N-1} &= h_2^* \\
      &\vdots \\
    h_{\frac{N}{2}+1} &= h_\frac{N}{2}^* \qquad \textrm{even } N \\
    h_\frac{N+1}{2} &= h_\frac{N+1}{2}^* \qquad \textrm{odd } N 
\end{eqnarray*}
\end{itemize}
It is easy  to check that equations \eref{ec:bpar} and \eref{ec:bimpar} remain {\it real } under these
identifications which make
obviously the periodic potential {\it complex} but $\mathcal{PT}$-invariant.
There has been an earlier
attempt to generate {\it real} band condition from a complex but $\mathcal{PT}$-invariant potential
\cite{BENDUNNE}. However the results concerning the appearance and
disappearance of forbidden
and allowed bands was inconclusive. In our case this effect is clear and
will be discussed at length
below.

As is well known for years a hermitian periodic potential cannot alter its
band spectrum just by fine tunning
the couplings. The bands can indeed be made wider or narrower but its
number and quality (forbidden or
allowed) remains unchanged. The theorems supporting these statements are
all based upon the intuitive idea
that a hermitian operator cannot change its spectrum that is basically
given by the eigenvalues and
eigenfunctions of the states at the edges of the bands. The mathematics can
be hard but the physical idea
was indeed whether this behaviour would be mantained if a $\mathcal{PT}$-invariant potential is used. For this purpose
they use various analytical potentials carefully shifted to be $\mathcal{PT}$-invariant. They can prove that the
band condition is real but in order to analyze the band structure they have
to assert with a very high degree
of precision whether a given curve is above (below) $+1$ ($-1$) in a
similar manner as we have to ascertain
ourselves that the expressions \eref{ec:cos1}-\eref{ec:cos3} (and in
general \eref{ec:bpar} and \eref{ec:bimpar}) exceeds $+1$ or
goes below $-1$. In the case of
\cite{BENDUNNE} this appears a very hard task indeed as the authors do not
have to their disposal an
analytical band condition, so they must carry out various kinds of
approximations. The authors conclude that
"{\it despite this impressive precision, [...its equations] (16) and (17)
cannot be used directly to answer
the crucial question of whether there are band gaps because these
approximations to the discriminant [... band
condition] never cross the values [...normalized to] $\pm 1$}"

But we do have such an exact band condition and in spite of the
apparent formidable aspect of the expressions
\eref{ec:bpar}-\eref{ec:bimpar} we can perform various kinds of exact and numerical
analysis in order to check the dependence of the  band width and the band number on the exceptional parameter that
arises in our model: \textbf{the imaginary
part of the complex couplings}.  This will be the subject of next Section.
%%%%%%%%%%%%%%%%%%%%%%%%%%%%%%%%%%%%%%%%%%%%%%%%%%%%%%%%%%%%%%%%%%%%%%%%%%%%%%%%
\section{The band condition for couplings with non-vanishing imaginary part}%
\label{sec:spectrum}
    In order to understand the effect of the imaginary part of the couplings on
the band spectrum one must analize in detail the behaviour of the band
condition. Let us begin with the simplest \PTs chain including $2$ deltas
in the primitive cell characterized by the parameters  $(a/a_1)=r_1+\rmi
s_1$ and $(a/a_1)^*=r_1-\rmi s_1$. This case has been also carefully studied considering different
distances between the deltas in reference \cite{AHMED}. The
band structure is entirely determined by the function
$\mathcal{B}=2\;h_1h_1^*-1$, which can written as
\begin{equation}
    \label{eq:extra}
    \mathcal{B}=\mathcal{B}(s_1=0) + \frac{2 s_1^2 \sin^2(\varepsilon)}{\varepsilon^2}.
\end{equation}
The term due to the imaginary part is always positive and it
has the effect of  lifting up the band condition for all values of $r_1$. 
The position and width of the allowed (forbidden) bands come from the
intersections at $\mathcal{B}=1,-1$ which depend strongly on
the position of the maxima and minima of the band condition. The
changes on these limiting points can be analytically described in this case. 
\begin{figure}
    \subfigure[$r_1=5$,
$s_1=4$]{\epsfig{file=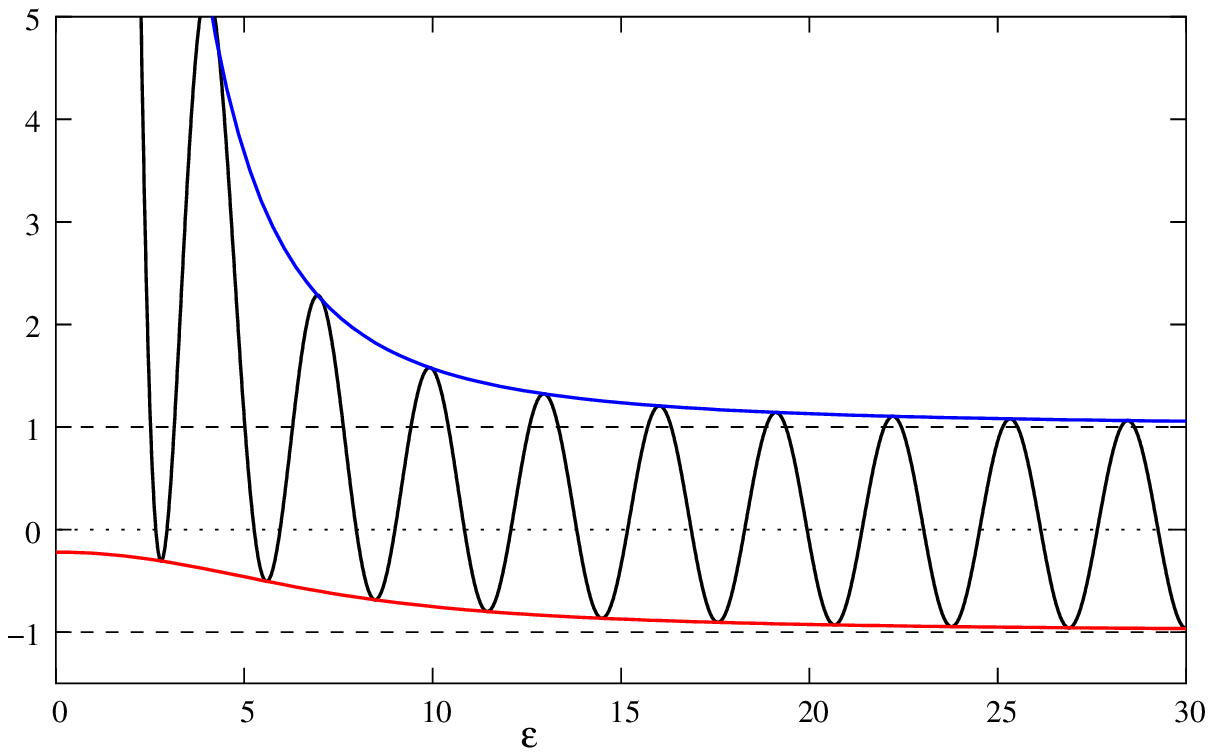,width=.49\textwidth}}
    \subfigure[$r_1=0.5$, $s_1=15$]{\epsfig{file=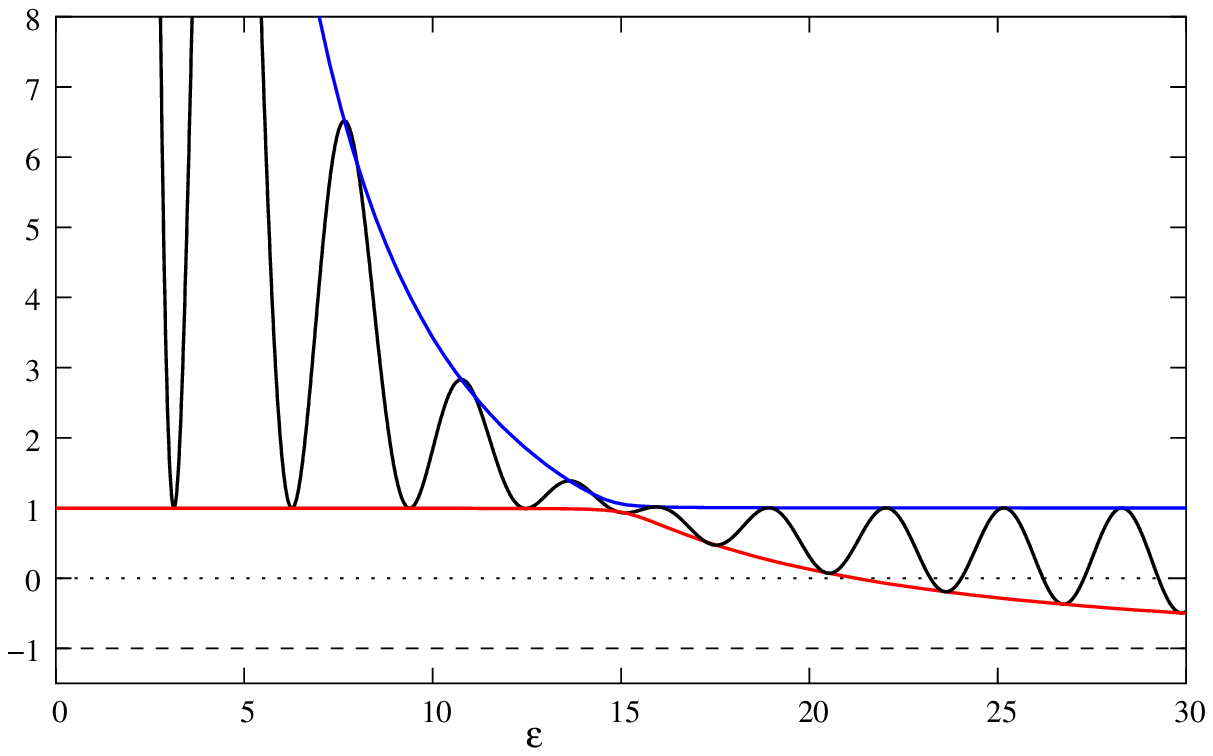,width=.49\textwidth}}
    \caption{Band condition for $N=2$ when $s\sim r$ and $s\gg r$. The
coloured lines corresponds to $C_\pm$.}
    \label{fig:bc2}
\end{figure}

From the equation $\rmd \mathcal{B}/\rmd \varepsilon=0$ one can find the
value of the oscillatory part of the band condition (i.e. the trigonometric
functions) as a function of $\varepsilon$ and substituting into $\mathcal{B}$
with a properly choice of signs will result in an analytical form for the
curves crossing the extremal points of the band condition which reads: 
\begin{equation}
    C_{\pm}=2\;f_{\pm}f_{\pm}^*-1
\end{equation}
where $C_+(C_-)$ lays on the maxima(minima) and the functions $f_\pm$ are
defined as follows:
\begin{eqnarray}
       f_{\pm} =& \frac{1}{\sqrt{F_\pm}}\left[\sqrt{2} r_1 (r_1-\rmi s_1) \pm
    \left(F_\pm-2 r_1^2 \varepsilon^2\right)^{\frac{1}{2}}\right] \\
    F_{\pm} =& r_1^2+(r_1^2+s_1^2)^2+2(r_1+r_1^2-s_1^2)\varepsilon^2
    +\varepsilon^4 \pm \nonumber\\
    &(r_1-r_1^2-s_1^2+\varepsilon^2)\left[(r_1+r_1^2+s_1^2)^2+2(r_1+r_1^2-s_1^2) \varepsilon^2+\varepsilon^4\right]^{\frac{1}{2}}.
\end{eqnarray}
Hence, the behaviour of the band condition can be studied through the evolution
of $C_\pm$. Taking the limit $s_1 \rightarrow 0$ we obtain,
\begin{equation}
    \lim_{s_1 \rightarrow 0} C_\pm = \cases{
    1+\frac{2r_1^2}{\varepsilon^2}\frac{r_1(2+r_1)+\varepsilon^2}{r_1(2+r_1)+\varepsilon^2+\frac{r_1^2}{\varepsilon^2}}\\
    -1.}
\end{equation}
The real part of the couplings controls the amplitude of the oscillations
(and therefore the distance between $C_+$ and $C_-$)
but the minima will always lean on $\mathcal{B}=-1$ unless the imaginary
part is non zero. Thus as we increase $s_1$ the minima will rise,
as can be seen in figure \ref{fig:bc2}(a), and they never
touch $-1$ because $\lim_{\varepsilon\rightarrow\infty} C_- = -1$. On the other hand considering $r_1\rightarrow 0$, after
some algebra one obtains
 \begin{eqnarray}
    \lim_{r_1 \rightarrow 0} C_+ &= \cases{
    -1+\frac{2s_1^2}{\varepsilon^2}\frac{s_1^4+\varepsilon^4+s_1^2(1-2\varepsilon^2)}                   
        {s_1^4+\varepsilon^4+s_1^2(\frac{s_1^2}{\varepsilon^2}-2\varepsilon^2)} &
    for $\varepsilon\leq s_1$\\ 1 & for $\varepsilon > s_1$}
\end{eqnarray}
and for $C_-$ the limit is the same as for $C_+$ but interchanging with respect
to the intervals in $\varepsilon$. Thus for an imaginary part of the coupling large enough
compared to the real one the band condition goes above $+1$ for energies
below the value $s_1$. Figure \ref{fig:bc2}(b) shows this last
configuration. In fact the situation $s_1\gg r_1$ can be easily understood
from the form of the band condition that can easily be written in this limit as:
\begin{equation}
    \mathcal{B}\simeq -1
+\frac{2s_1^2}{\varepsilon^2}+2\cos^2(x)\left(1-\frac{s_1^2}{\varepsilon^2}\right)
\end{equation}
which shows clearly the boundaries when $\varepsilon \leq
s_1:\,1\leq\mathcal{B}\leq -1+2s_1^2/\varepsilon^2$ saturated at
$\varepsilon=n\pi$ and $\varepsilon=(2n+1)\frac{\pi}{2}\quad n\in \mathbb{Z}$
respectively. For $\varepsilon > s_1$ the boundaries simply interchange
among them. Due to the
nature of the functions $h_j(\varepsilon)$ the band condition is always tied up to
the value $+1$ at every multiple of $\pi$ for all $r_1, s_1$, as
figure \ref{fig:bc2}(b) clearly shows. Several configurations of the spectrum as the
ones in figure \ref{fig:bc2} can be
built for different values of the complex coupling.

\begin{figure}
    \subfigure[$r_1=4$, $r_2=5$, $s_1=3$]{\epsfig{file=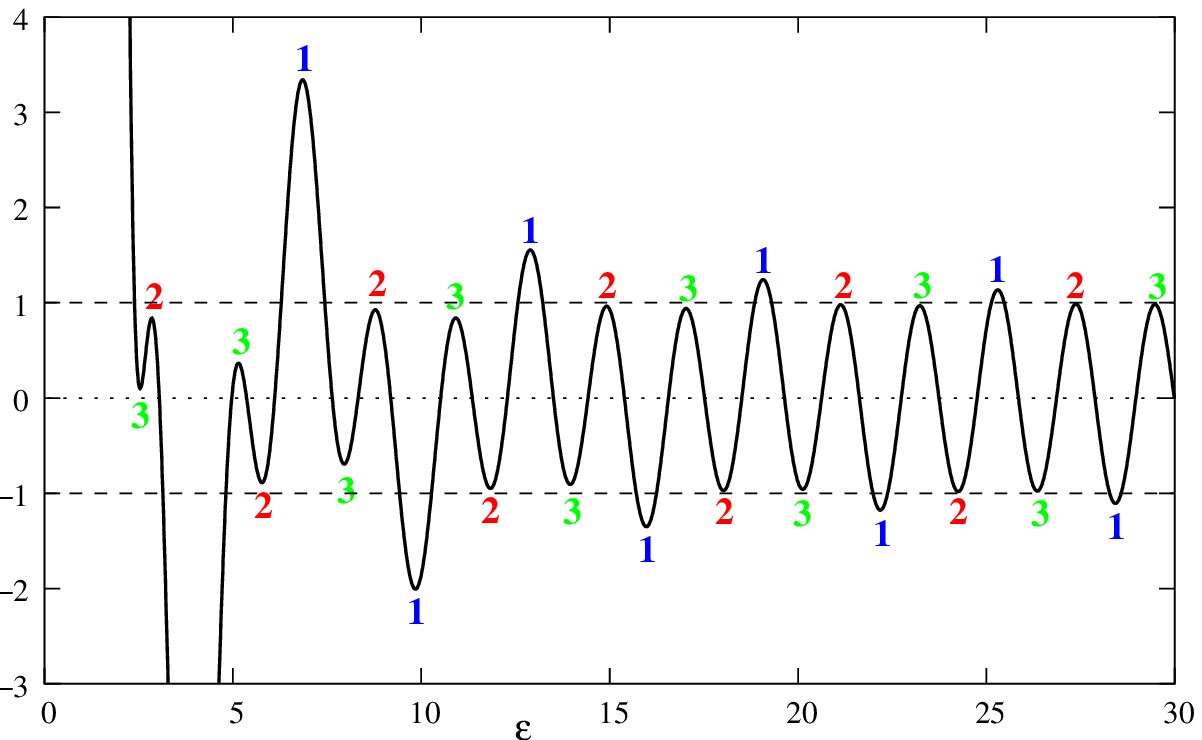,width=.49\textwidth}}
    \subfigure[$r_1=3$, $r_2=2$, $s_1=5$]{\epsfig{file=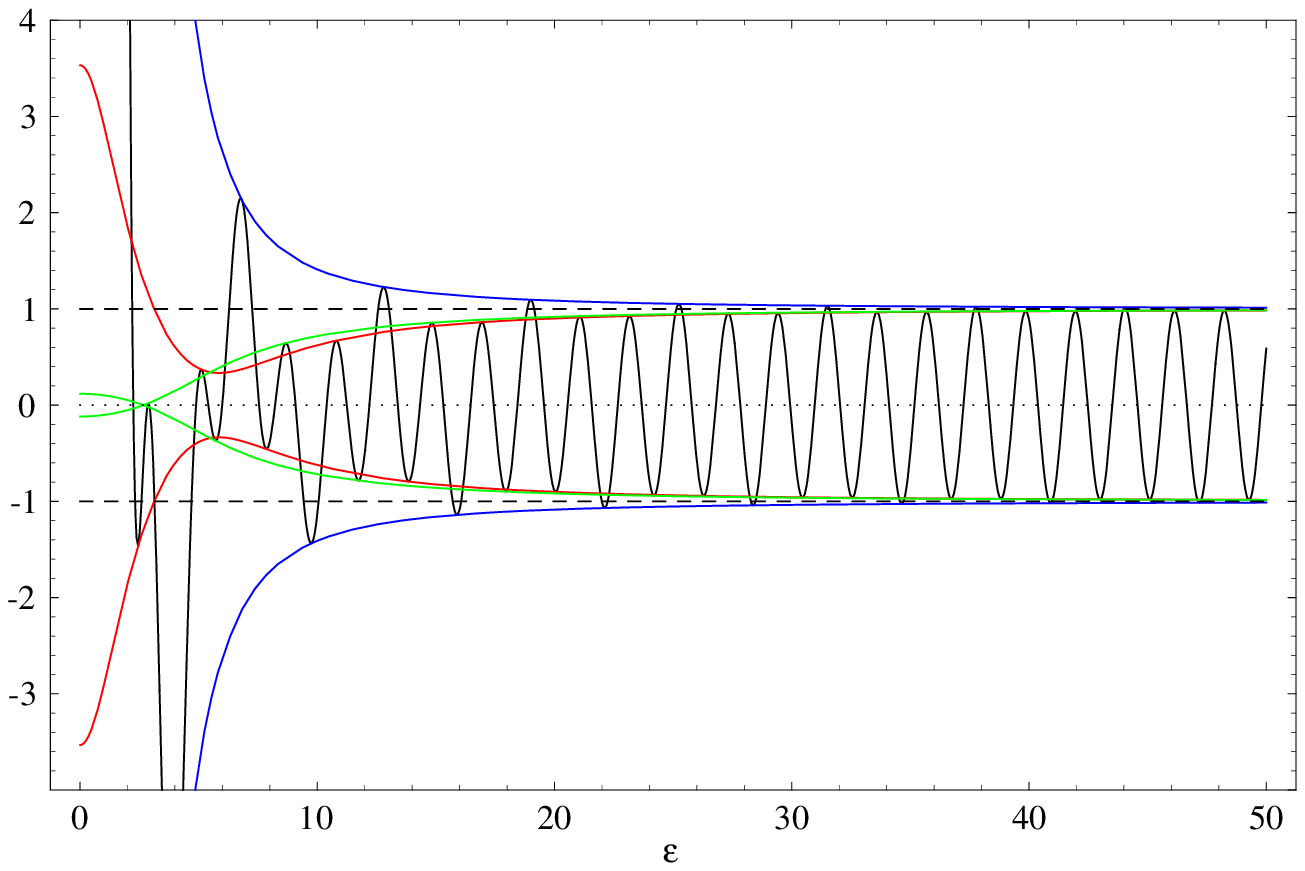,width=.49\textwidth}}
    \subfigure[$r_1=3$, $r_2=2$, $s_1=25$]{\epsfig{file=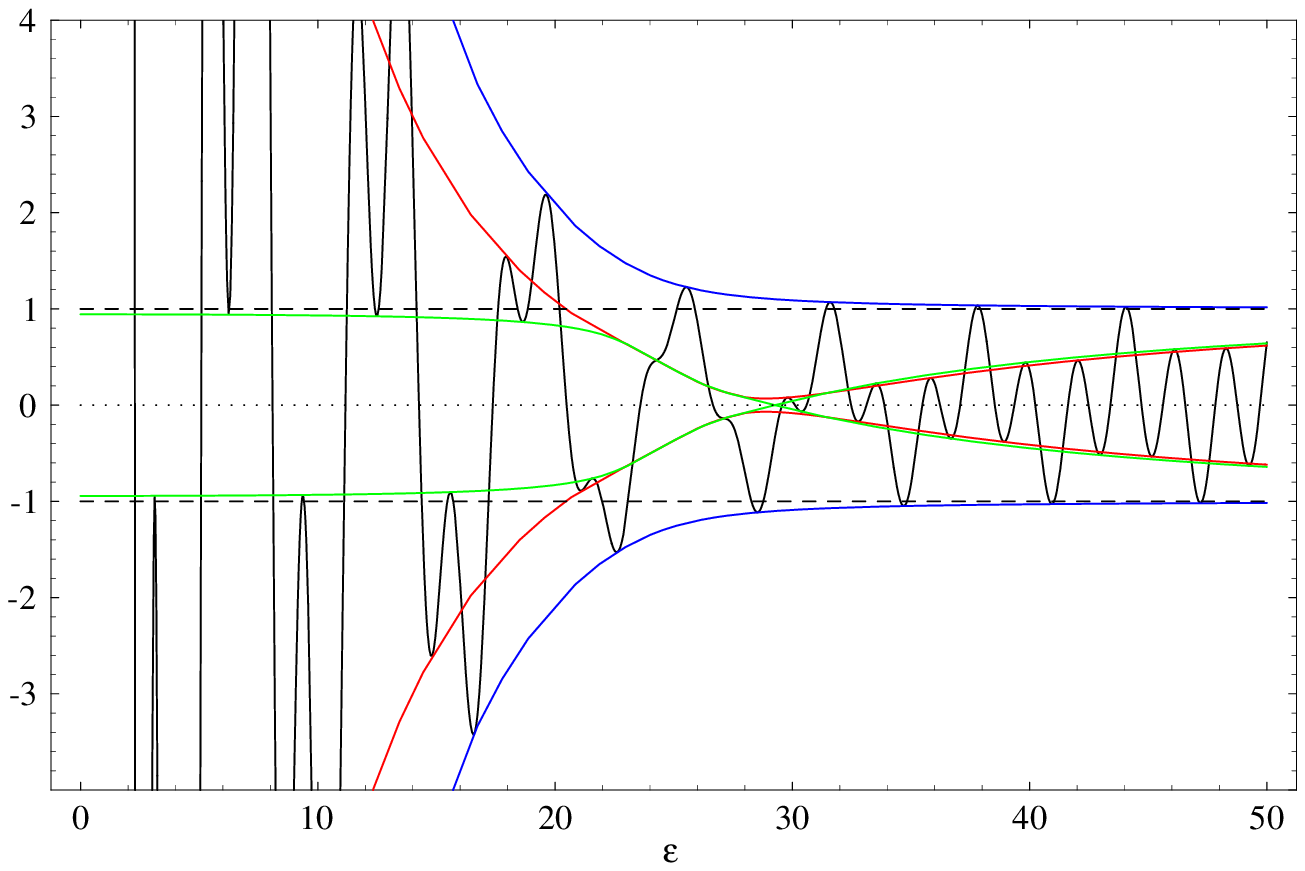,width=.49\textwidth}}
    \subfigure[$r_1=10$, $r_2=10$, $s_1=25$]{\epsfig{file=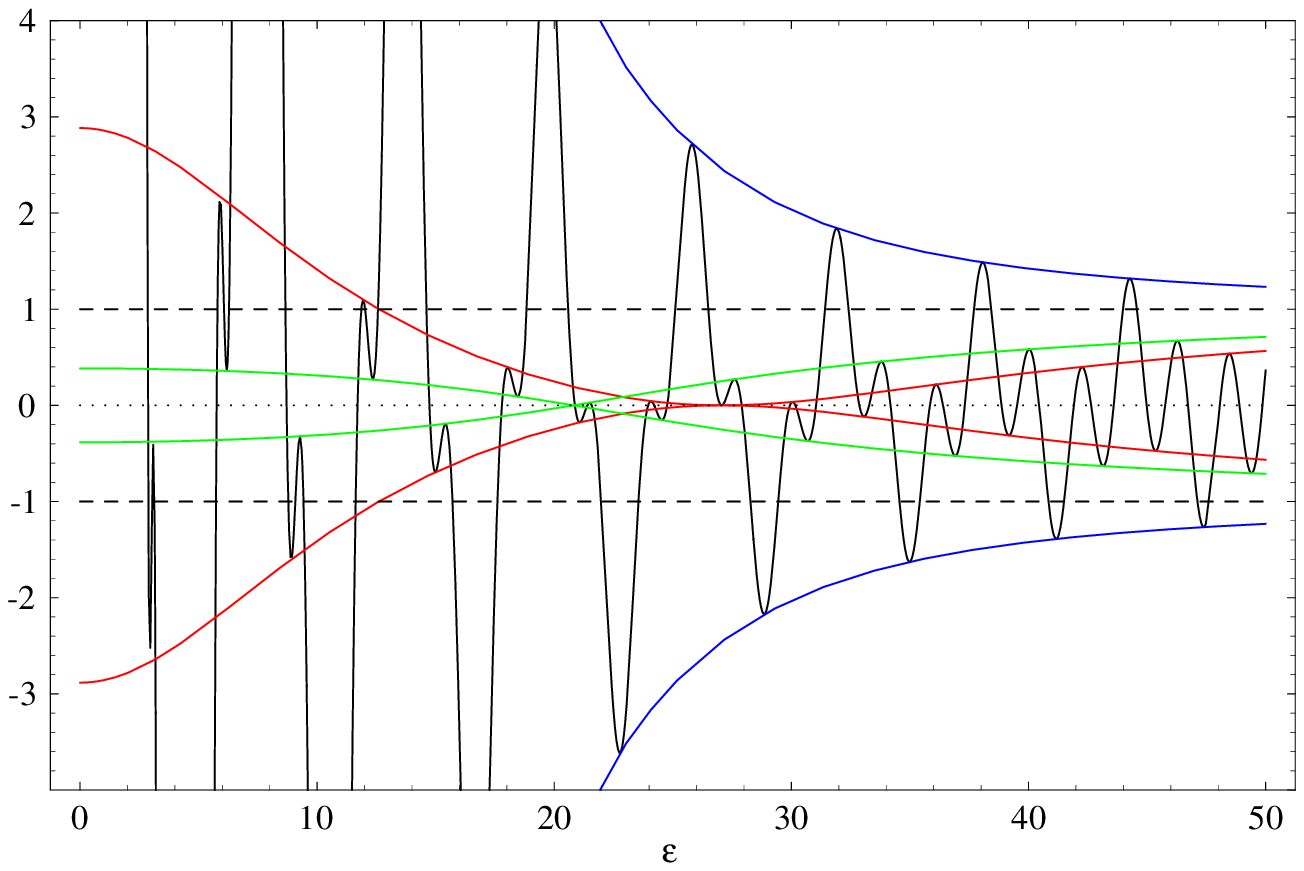,width=.49\textwidth}}    
    \caption{Band condition for $N=3$ for different values of the
couplings. Each type of extremal point is labelled with a different colour.}
    \label{fig:bc3}
\end{figure}
Let us now consider a \PTs chain whose primitive cell includes $3$ atoms
charaterized by the parameters $(a/a_1)=r_1+\rmi s_1$, $(a/a_2)=r_2$ and
$(a/a_1)^*$. The
band condition function reads $\mathcal{B}=4\;
h_1h_1^*h_2-(h_1+h_1^*+h_2)$. This can be arranged as:
\begin{equation}
    \label{ec:bc3}
    \mathcal{B}=\mathcal{B}(s_1=0)+\frac{4 s_1^2 \sin^2(\varepsilon)
\left[\varepsilon \cos(\varepsilon) + r_2 \sin(\varepsilon)\right]}{\varepsilon^3}.
\end{equation}
The extra term arising from the imaginary part of the coupling can be
either positive or negative depending on the value of $r_2$ and the
energy. The effect of $s_1$ is different compared to the previous case
although there is only one imaginary parameter.
The structure of the band condition seems to be composed by 
pieces each one including $3$ different types of extremal points, as figure
\ref{fig:bc3}(a) shows, and each type
apparently following a given pattern. The
curves crossing the extremal points of the band condition can also be analytically
calculated in this $N=3$ case but the expressions are quite hard to
simplify and therefore we do not provide the explicit form of the equations. Three curves are obtained, one for each type
of extremal point. Initially when $s_1=0$ all the extremal points are outside or on the
borders of the range $[-1,1]$. As we increase the imaginary part, the
amplitudes of two of the three types
of extremal points begin to decrease (green and red in figure
\ref{fig:bc3}(b)). At the same time as $s_1$
becomes larger the green and red curves get narrower with decreasing energy 
 until $\varepsilon \sim s_1$. Below this value the curves broaden
trying to expel the extremal points of the band condition outside the
target range (figure \ref{fig:bc3}(c)). From the form of $\mathcal{B}$ it is
clear that the even(odd)
multiples of $\pi$ will remain fixed to $1$($-1$). The efficiency of this expelling process depends upon the values of
the real couplings as they control the amplitudes of the oscillations. One
can stretch the maxima and minima (up and down respectively) increasing
$r_1$, $r_2$, as shown in figure \ref{fig:bc3}(d).

Several configurations can be obtained among the ones shown in figure
\ref{fig:bc3} and one important feature must be emphasized: the band
condition always changes ``symetrically'' with respect to the abscisa
axis. 
 That is: there is the same amount of positive and negative function whatever
the values of the couplings are. In fact the curves laying on the extremal points
show an exact reflection symmetry around the energy axis. This is in great contrast to the $N=2$
case where the band condition can be pushed up above $+1$ for certain
couplings. Thus when $s_1\gg r_1,r_2$ there will always remain a trace of the original permitted
energy ranges in the form of  ``flat'' bands.  

\begin{figure}
    \subfigure[$r_1=3$, $r_2=2$, $s_1=0$, $s_2=0$]{\epsfig{file=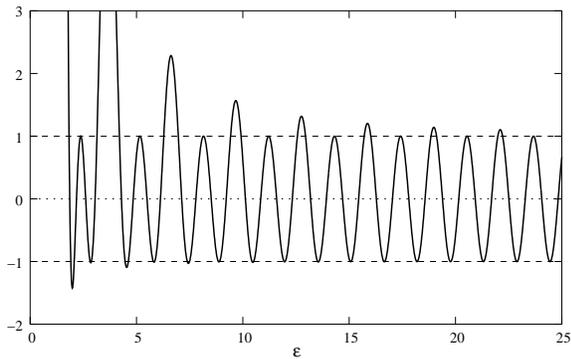,width=.49\textwidth}}
    \subfigure[$r_1=3$, $r_2=2$, $s_1=9$, $s_2=2$]{\epsfig{file=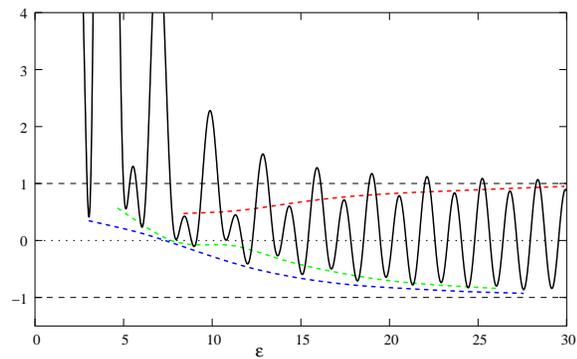,width=.49\textwidth}}
    \subfigure[$r_1=3$, $r_2=2$, $s_1=4$, $s_2=25$]{\epsfig{file=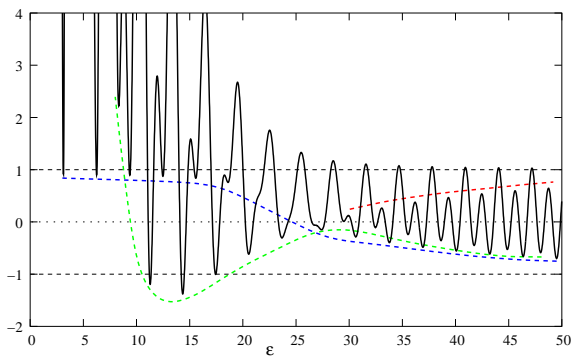,width=.49\textwidth}}
    \subfigure[$r_1=3$, $r_2=2$, $s_1=20$, $s_2=21$]{\epsfig{file=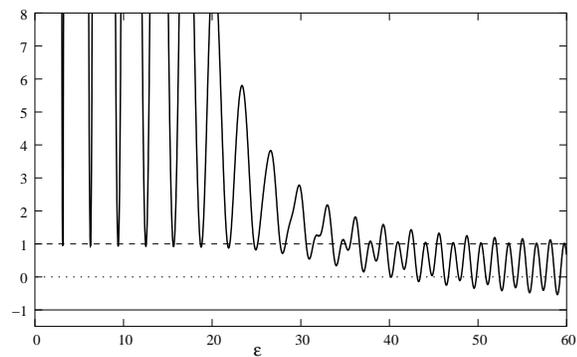,width=.49\textwidth}}    
    \caption{Band condition for $N=4$ with different values of the
couplings. The coloured dashed lines are intended to clarify the behaviour of
different groups of extremal points.}
    \label{fig:bc4}
\end{figure}
 The first chain for which two different imaginary couplings can be
manipulated is the one with $N=4$ atoms inside the primitive cell:
$(a/a_1)=r_1+\rmi s_1$, $(a/a_2)=r_2+\rmi s_2$, $(a/a_2)^*$ and
$(a/a_1)^*$. In this case we have not calculated analytically the curves
crossing the extremal points.  
Nevertheless a pictorial approach for
some of the curves has been included in the graphics for a better understanding of
the tendencies of change. The initial situation ($s_1=s_2=0$) is a common one for a
periodic chain (\fref{fig:bc4}(a)) and two different group of extremal points can
be distinguished. As we increase the imaginary part of the couplings one
group of extremal points starts to decrease and the band condition lifts
up globally (\fref{fig:bc4}(b)). The effect of the two imaginary parts is quite similar. When
both of them have comparable values the two groups of minima follow the
same tendency moving upwards the band condition. However if one imaginary
part grows much
more than the other one, several minima strech down in a region of energy
roughly included in $s_{small}<\varepsilon < s_{large}$
(\fref{fig:bc4}(c)). Finally when $s_1,s_2\gg r_1,r_2$  one can force the
band condition to go above $+1$ (keeping our well known knots at
$\varepsilon=n \pi$) in a certain energy range which depends on
the values of the imaginary parts of the couplings (\fref{fig:bc4}(d))
approximately the same way as for the $N=2$ case. Unlike the $N=3$ example the
band condition can be unbalanced to positive values with a properly choice
of the parameters.

\begin{figure}
    \subfigure[$r_1=4$, $r_2=1$, $r_3=1$, $s_1=20$, $s_2=5$]{\epsfig{file=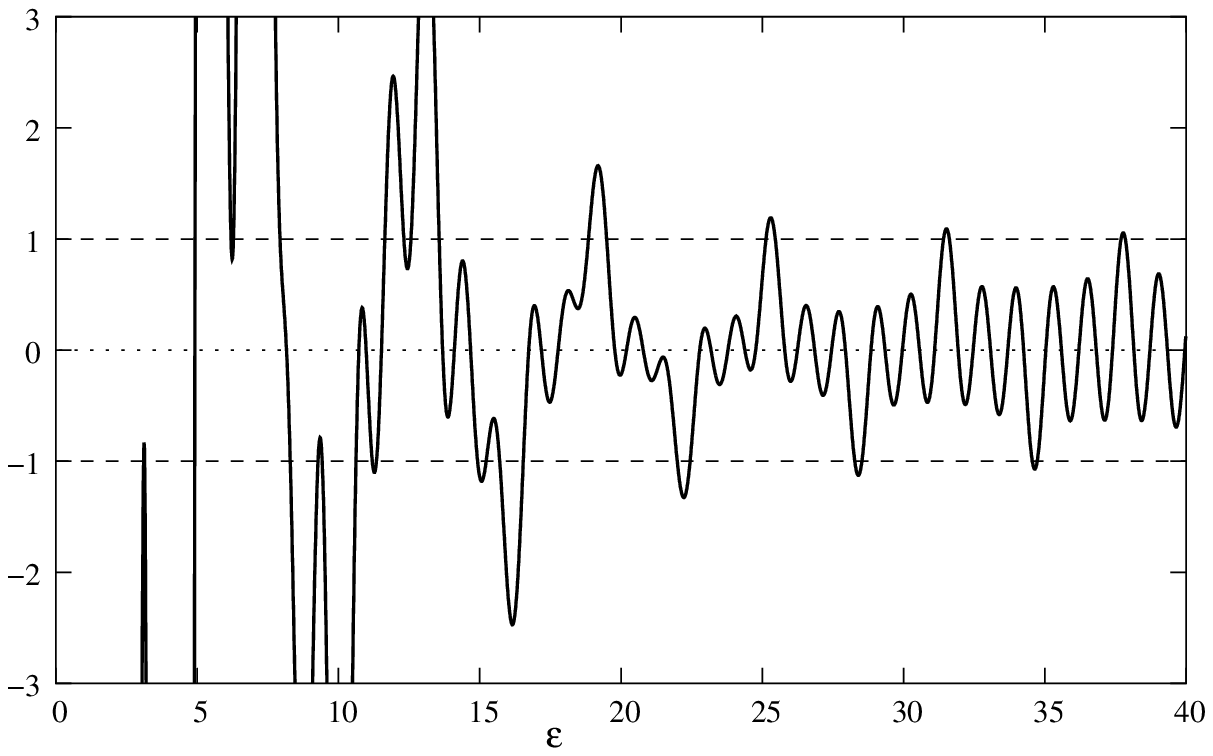,width=.49\textwidth}}
    \subfigure[$r_1=4$, $r_2=2$, $r_3=3$, $s_1=10$, $s_2=20$]{\epsfig{file=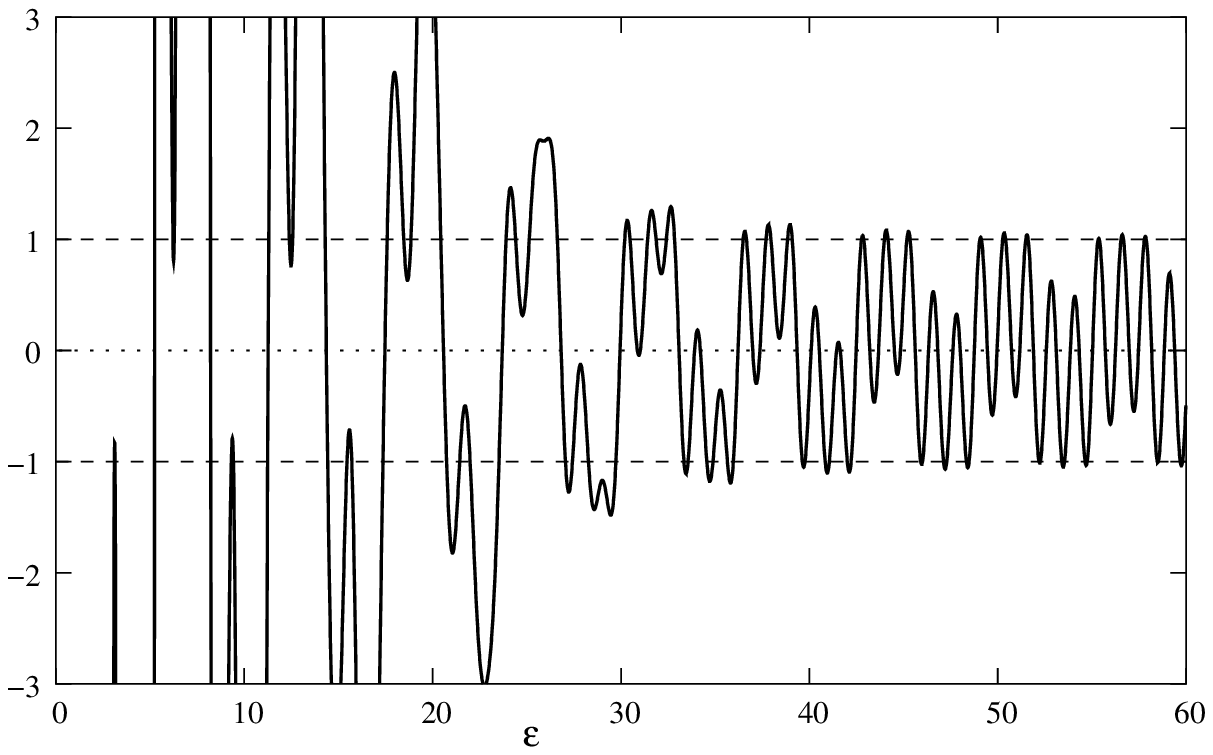,width=.49\textwidth}}
    \caption{Band condition for a \PTs chain with $N=5$.}
    \label{fig:bc5}
\end{figure}

\begin{figure}
    \subfigure[$r_1=5$, $r_2=3$, $r_3=4$, $s_1=2$, $s_2=5$, $s_3=4$]{\epsfig{file=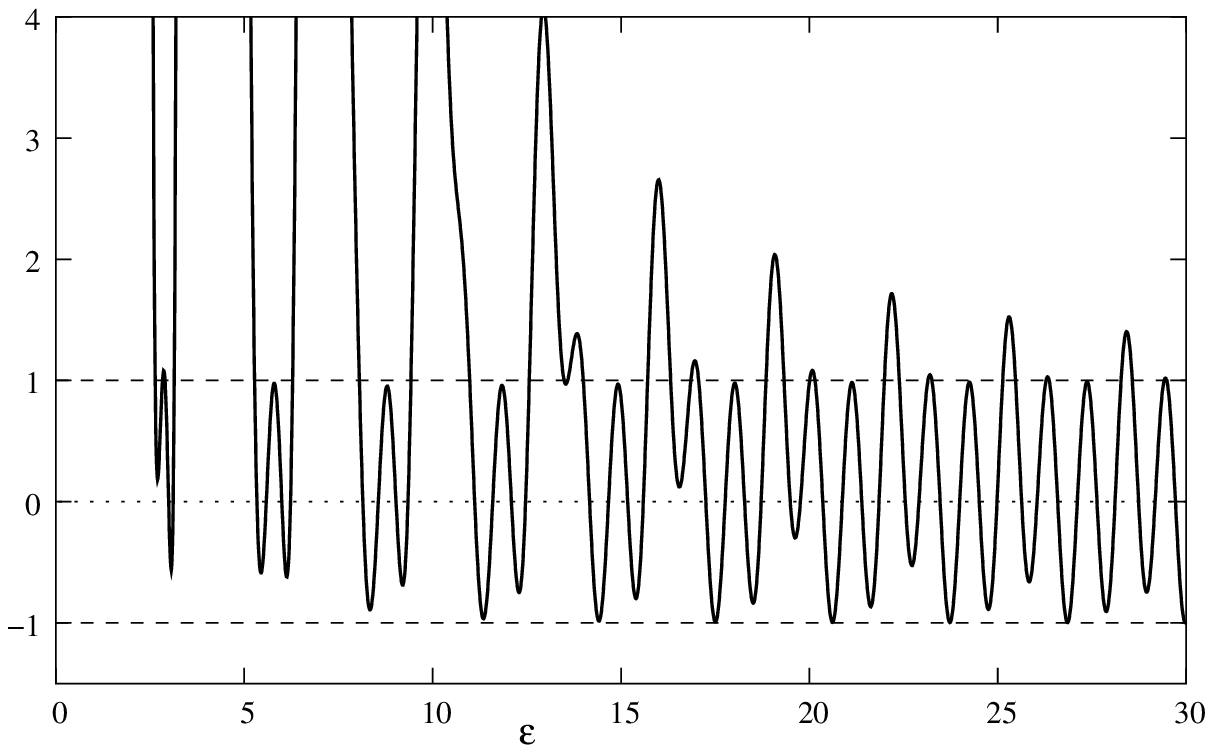,width=.49\textwidth}}
    \subfigure[$r_1=1$, $r_2=2$, $r_3=0.5$, $s_1=10$, $s_2=11$, $s_3=12$]{\epsfig{file=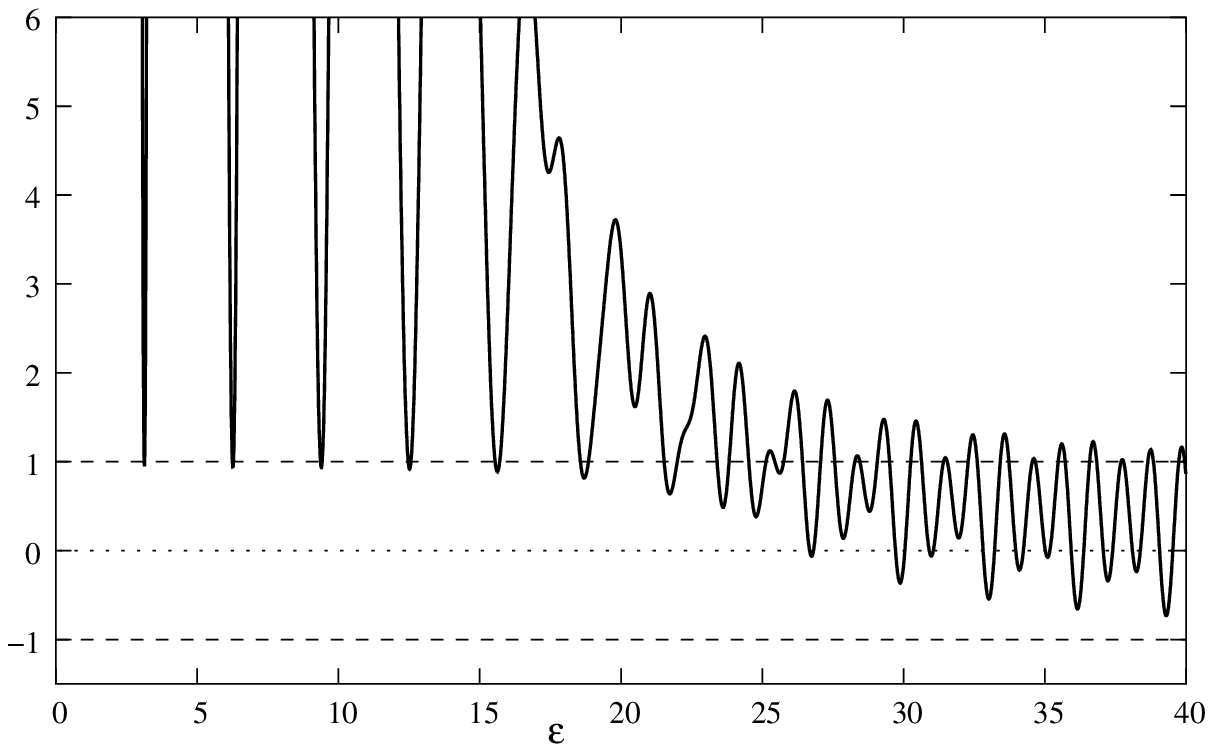,width=.49\textwidth}}
    \caption{Band condition for a \PTs chain with $N=6$.}
    \label{fig:bc6}
\end{figure}

The band condition has been also studied in detail for $N=5$ and
$N=6$. Its behaviour gets really complex as the number of atoms grows. In 
\fref{fig:bc5} and \fref{fig:bc6} some characteristic examples
are shown. For $N=5$ the band condition shows five types of extremal points that
evolve ``symmetrically'' around the abcisa axis as we change the imaginary
couplings. For $N=6$ three types of maxima and minima appear and for large
enough values of the imaginary couplings the band condition is greater than
$+1$.

 To summarize this mathematical analysis of the band
condition let us now list some of the most salient features:
\begin{itemize}
    \item the band condition is composed of pieces integrated by $N$($N/2$)
extremal points for odd(even) $N$. Each extremal point belongs to a group that evolves
differently according to the values of the imaginary parts of the
couplings
    \item for odd $N$ the band condition is ``symmetric'' around the
abcisa axis for all values of the couplings. Therefore, some permitted
levels always remain as a part of the spectrum
    \item for even $N$ the band condition can be expelled out of the target
range for certain values of the couplings and so removing the allowed bands
    \item for $\varepsilon=n\pi$ the band condition is fixed to $+1$ or $-1$
depending on the parity of $N$ and $n$ for all values of the couplings
    \item the real parts of the couplings are always proportional to the amplitude of
the oscillations. 
\end{itemize}
Even with all the included figures it is hard to make a complete explanation
of the behaviour of the band condition that otherwise can only be fully 
understood watching some proper animations of the function. This is the
procedure we have followed to support the results. We encourage the
interested reader to reproduce those animations that can easily be done
with \texttt{Mathematica}. Some example code lines are listed
in the appendix.
%%%%%%%%%%%%%%%%%%%%%%%%%%%%%%%%%%%%%%%%%%%%%%%%%%%%%%%%%%%%%%%%%%%%%%%%%%%%%%%%%%%%%%
\section{The band structure and the electronic states}
\label{sec:physics} 
\begin{figure}
     \subfigure[]{\epsfig{file=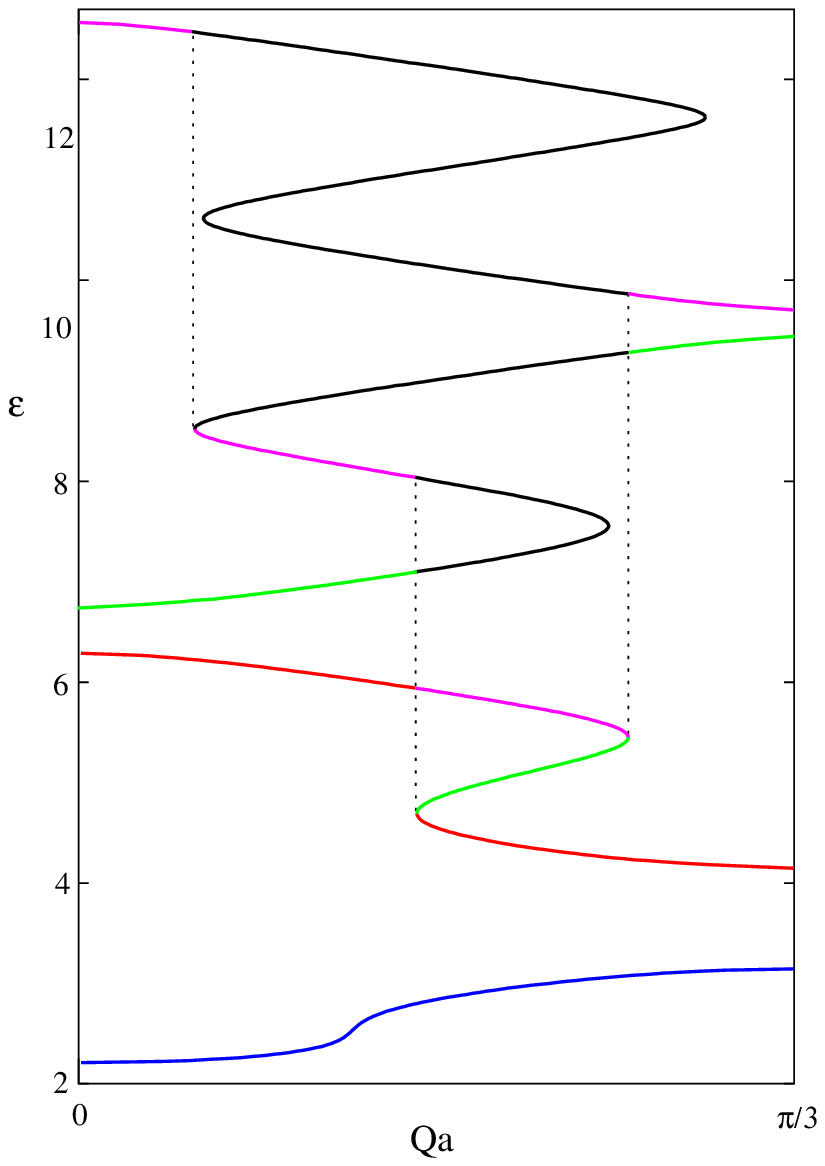,width=.32\textwidth}}
    \subfigure[]{\epsfig{file=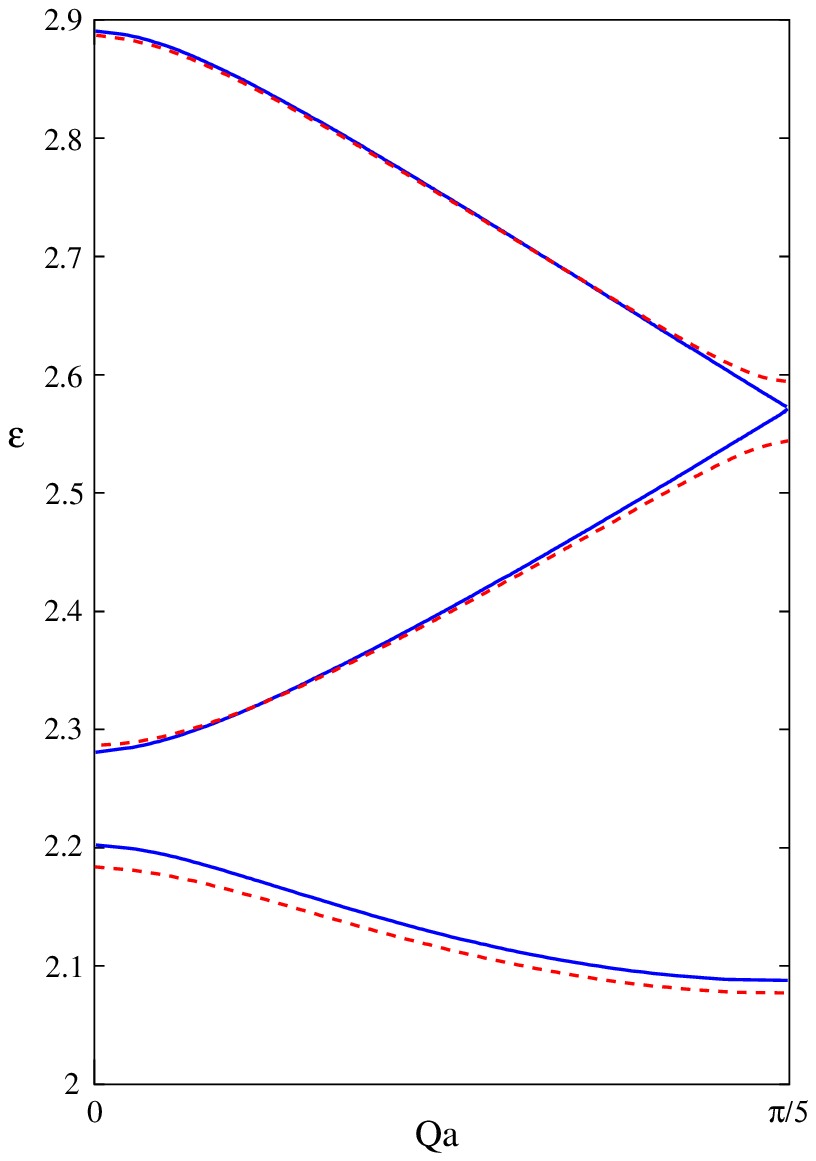,width=.32\textwidth}}
     \subfigure[]{\epsfig{file=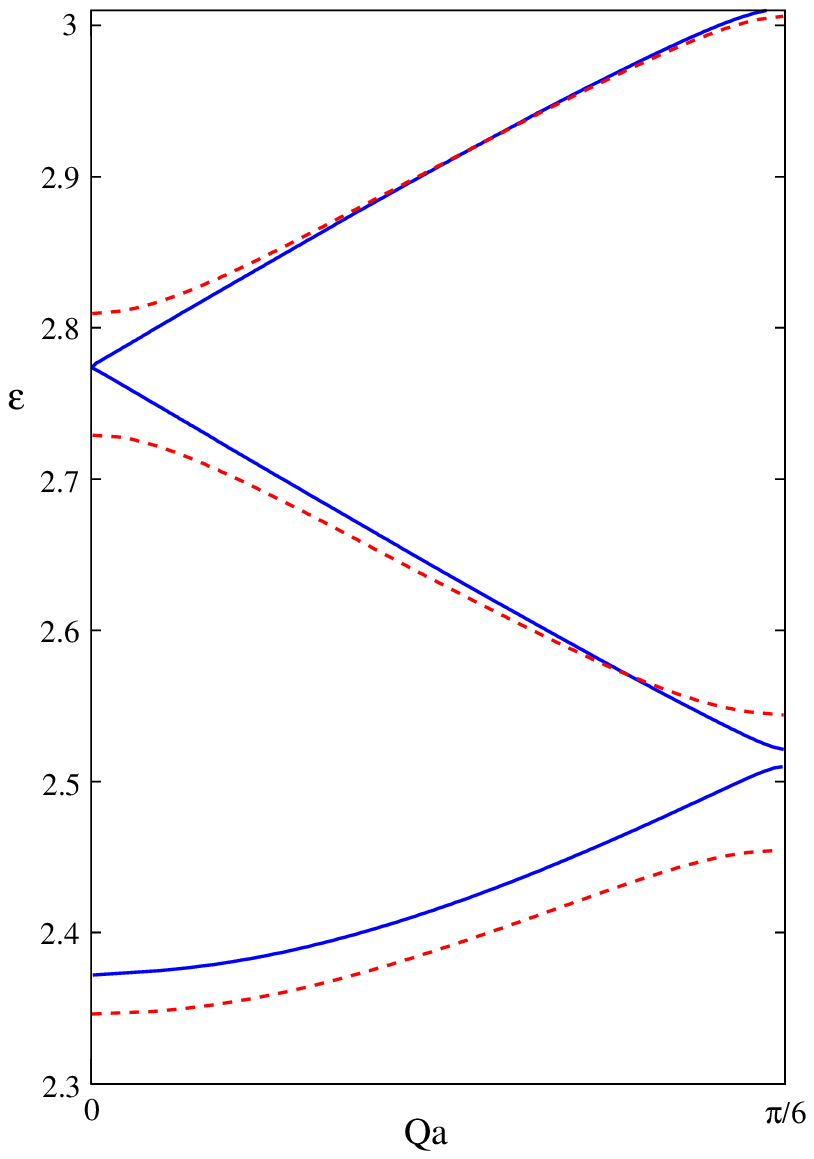,width=.32\textwidth}}
    \caption{Band structure in the irreducible part of the first Brillouin
zone for different chains: \textbf{(a)} $N=3$ with $r_1=1$, $r_2=2$,
$s_1=3$. Different colours show the first four bands.
 \textbf{(b)} $N=5$ with $r_1=4$, $r_2=2$,
$r_3=1$, $s_1=s_2=0$ (red) and $s_1=0.775$, $s_2=0.1$ (blue). \textbf{(c)}
$N=6$ with $r_1=5$, $r_2=2$, $r_3=3$, $s_1=s_2=s_3=0$ (red) and
$s_1=1.613$, $s_2=0.12$, $s_3=0.3$ (blue).}
    \label{fig:bands}
\end{figure}

Let us now make some comments about the band structure.
One important feature that stands out from the above examples is the
presence of maxima(minima) inside the target range $(-1,1)$. This fact
involves the presence of points in the recripocal space where the gradient
of the energy diverges that makes the physical interpretation
of the hamiltonian not so straightforward. Probably some additional restrictions have to be imposed on
the \PTs hamiltonian. These special points also cause
some changes on the form of the bands which are always understood as the ensembles of states
and energies labelled with a certain index ($n$ for the $n$-th band). Every point of the
Brillouin zone must represent a physical state on every band. Therefore for a certain
value $Qa$ let us label the eigenenergies of the hamiltonian following an
increasing order ($\varepsilon_1(Qa) < \varepsilon_2(Qa) < \ldots$). These procedure
leads unavoidably to the definition of bands whose dipersion relations are
not continuos over the reciprocal space as seen in \fref{fig:bands}(a). Of course the choice of the bands indices is
not unique but none of them will be free of these discontinuities.

On the other hand those pathologies can be avoided if one restrics his
working scenario to an small range of energies. In that case the imaginary
parts of the couplings can be tuned to modify essentially the spectrum of
the system removing or decreasing gaps to change virtually its response to transport
phenomena (figures \ref{fig:bands}(b) and \ref{fig:bands}(c)).

\begin{figure}  
    \centering
    \epsfig{file=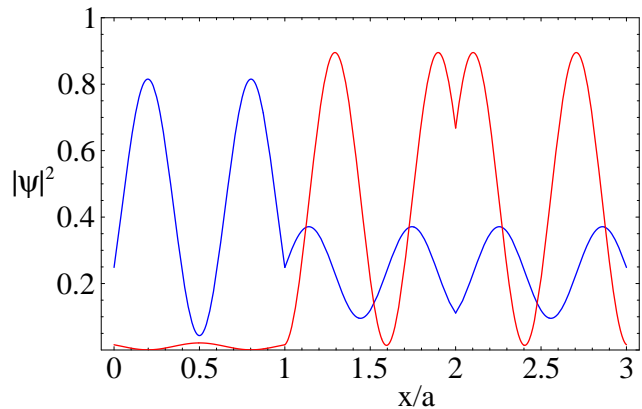,width=.6\textwidth}
    \caption{Electronic wave function normalized to 1 inside the primitive
cell of a \PTs chain for $N=3$ with $r_1=5$, $r_2=3$. $s_1=0$(blue) and
$s_1=19$(red) at $\varepsilon=5.2$.}
     \label{fig:wave1}
\end{figure}
\begin{figure}
    \centering
    \epsfig{file=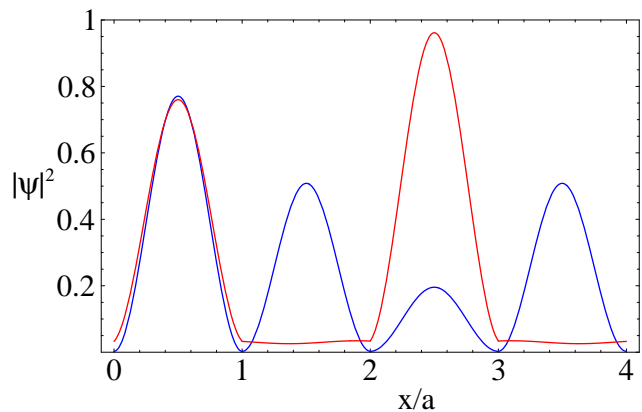,width=.6\textwidth}
    \caption{Electronic wave function normalized to 1 inside the primitive
cell of a \PTs chain for $N=4$ with $r_1=3$, $r_2=4$, $s_1=5$, $s_2=8.1$
for different energies: $\varepsilon=2.95$(red)
and $\varepsilon=3.13$(blue).}
    \label{fig:wave2}
\end{figure}
One can also wonder about the form of an electronic states belonging to such a
characteristic spectrum. It is not hard to calculate analytically the wave
function inside the primitive cell as a function of the position and the
energy with the help of the computer for low $N$. To our surprise we have
found that the imaginary parts of the couplings behave as control parameters
of the localization of the electrons inside the primitive cell. Thus one could
tune these parameters to decrease the probability of presence in several
sectors almost to zero or distribute it more homogeneously over the
primitive cell, as shown in \fref{fig:wave1}. Also for non-vanishing
imaginary parts of the couplings the depence of the state on the energy
seems quite strong as a small variation of this energy can involve an important
change in the shape of the state (\fref{fig:wave2}).
%%%%%%%%%%%%%%%%%%%%%%%%%%%%%%%%%%%%%%%%%%%%%%%%%%%%%%%%%%%%%%%%%%%%%%%%%%%%%%%%%%
\section{Conclusions}
In this paper the band structure of a one dimensional $N$-different
coupling delta periodic potential is also assumed to be \PTs
and hence the delta couplings can be made complex. In spite of the
non-hermitian nature of such a Hamiltonian, the band spectrum condition
remains real. The imaginary parts of the couplings play an essential role
in the variety of shapes and forms that the band spectrum and the
wavefunctions may acquire in this new framework. Although the band
condition is analytical for arbitrary $N$, we have studied carefully just
some cases with a low number of couplings. A throughout study of the $N =
2$ case is given with analytic expressions for the limiting points of the
bands. The cases $N=3$, $4$, $5$ and $6$ are solved graphically and conclusive
results for the band spectrum are also offered.  In all these cases, the control
of the imaginary parts plays an essential role in the way that the band
spectrum gets distorted and reshaped. As we have been able to discuss the
evolution of this band spectrum, we are now planning to apply these results
to more physical situations in which localization and de-localization may
be tuned on and off by means of a judicious control of the imaginary parts.
We have been able to show how wavefunction de-localization takes place
actually for particular cases as the ones discussed in Section
\ref{sec:physics}. However a
manifold of interest questions remains to be answered such as: Could a
random (non-periodical) calculation with this sort of potentials be carried
out in the case of \PT?, Does the de-localization property
holds for more general cases?. These and related research subjects are now
being investigated and will be reported elsewhere.

%%%%%%%%%%%%%%%%%%%%%%%%%%%%%%%%%%%%%%%%%%%%%%%%%%%%%%%%%%%%%%%%%%%%%%%%%%%%%%%%
\ack
    We acknowledge with thanks the support provided by the Research in
Science and Technology Agency of the Spanish Government (DGICYT) under
contract BFM2002-02609.
%%%%%%%%%%%%%%%%%%%%%%%%%%%%%%%%%%%%%%%%%%%%%%%%%%%%%%%%%%%%%%%%%%%%%%%%%%%%%%%%
\appendix
\section*{Appendix}
    Here are some example lines for animating the band condition. First
load the package ``Animation'' and define the band conditions for $N=2$
and $N=3$.
{\tt
\begin{verbatim}
    << Graphics `Animation` 
    h[r_,s_]=Cos[x]+(r+I*s)*Sin[x]/x;
    B2[r1_,s1_]=2*h[r1,s1]*h[r1,-s1]-1;
    B3[r1_,r2_,s1_]=4*h[r1,s1]*h[r2,0]*h[r1,-s1]
                         -(h[r1,s1]+h[r2,0]+h[r1,-s1]);
\end{verbatim}}
\noindent To see the evolution of the band condition for $N=3$ versus the
imaginary part $s_1$ for $r_1=2$ and $r_2=3$ one would write:
{\tt \begin{verbatim}
    Animate[Plot[Evaluate[{B3[2,3,t],1,-1}],{x,0,30},   
                                         PlotRange->{-3,3}],{t,0,30}]
\end{verbatim}}
\noindent and finally one must group and compress the cells with all the generated plots
and by double clicking on the image the animation will start. 
%%%%%%%%%%%%%%%%%%%%%%%%%%%%%%%%%%%%%%%%%%%%%%%%%%%%%%%%%%%%%%%%%%%%%%%%%%%%%%
\section*{References}%

\end{document}